\documentclass[runningheads]{llncs}
\usepackage{amssymb}
\usepackage{graphicx}
\usepackage{subfigure}
\usepackage{url}
\urldef{\mailsa}\path|{thomas.silverston, olivier.fourmaux}@lip6.fr|
\newcommand{\keywords}[1]{\par\addvspace\baselineskip
\noindent\keywordname\enspace\ignorespaces#1}
\begin{document}
\mainmatter  
\title{P2P IPTV Measurement: A Comparison Study}
\titlerunning{P2P IPTV Measurement: A Comparison Study}
\author{Thomas Silverston\and Olivier Fourmaux}
\authorrunning{Thomas Silverston and Olivier Fourmaux}
\institute{Universit\'e Pierre et Marie Curie - Paris 6\\
104 avenue du Pr\'esident Kennedy, 75016 Paris, France\\
\mailsa\\
}
\toctitle{Lecture Notes in Computer Science}
\maketitle
\begin{abstract}
With the success of P2P file sharing, new emerging P2P applications arise on the Internet for streaming content like voice (VoIP) or live video (IPTV).
Nowadays, there are lots of works measuring P2P file sharing or P2P telephony systems, but there is still no comprehensive study about P2P IPTV, whereas it should be massively used in the future.
During the last FIFA world cup, we measured network traffic generated by P2P IPTV applications like PPlive, PPstream, TVants and Sopcast.
In this paper we analyze some of our results during the same games for the applications. 
We focus on traffic statistics and churn of peers within these P2P networks.
Our objectives are threefold: we point out the traffic generated to understand the impact they will have on the network, 
we try to infer the mechanisms of such applications and highlight differences, and we give some insights about the users' behavior.
\end{abstract}
\section{Introduction}
\label{section:introduction}
P2P video live streaming systems (P2P IPTV) are emerging applications, and should be massively used in the future. 
In these applications, end-users receive the flows and  duplicate them to other end-user peers.  
There is a lot of academic works that focus on the P2P networks architectures (\cite{infocom05:donet}, \cite{iptps06:chunkyspread}, \cite{infocom06:anysee}), but
there is still no particular attention to evaluate the traffic generated by P2P IPTV.
It is important for network managers and ISP to understand the traffic impact P2P IPTV will have on their network and the users' behavior of such applications.
\\
There are lots of works measuring traffic from P2P file sharing or P2P telephony system but there is no comprehensive measurement study about P2P IPTV.
This is the reason why we measured network traffic generated by several P2P IPTV applications during the last FIFA world cup\footnote{traces can be made available upon request}. 
Our main goals are to study the global behavior of such applications by observing the network and their users.
We also make comparisons between measured applications and try to infer general design.\\ 
We collected packet traces by using the following applications: PPLive, PPStream, SOPcast and TVants\footnote{http://www.pplive.com, http://www.ppstream.com, http://wwww.sopcast.com, http://www.tvants.com}. 
We chose these applications because they are the most popular on the Internet.\\
We measured generated traffic during the last FIFA World Cup because it is a large-scale event with live interest for users and we think this will allow us to have a representative sample of large-scale P2P IPTV applications.\\
Our works focus on P2P video live streaming applications (P2P IPTV). We classify P2P video streaming applications into two main categories:
video on-demand and live video. 
In the former one, the video is a pre-recorded finite file, which is sent to users whereas in the latter, the total size of the data is not known a priori and can be infinite.
Thus, these two different contents do not present the same interest for users since one is pre-recorded, and the other is not and broadcasted event is really occurring.
Sport events like Superbowl or FIFA soccer world cup present live interest compared to a pre-recorded movie.
We think measured traffic patterns would be different if the content exhibits live interest to users.
With the huge amount of collected data, we can study the impact of P2P IPTV applications on the network and we took advantage of the users' interest to watch this event in live to study their global behavior.
To the best of our knowledge, this is the first comprehensive measurement study about P2P IPTV systems during a large-scale event that exhibits a high interest to users.\\
In this paper, we aim to make comparisons between applications by analyzing the different traffic patterns we collected for all the applications.
It allows us to highlight design differences and to point out users' behavior in these P2P networks.
 In this work, we focus on a single event day where two soccer games were scheduled
and we analyze the traffic generated by the four previously mentionned P2P applications.\\
The remainder of this paper is organized as follows. In Section \ref{section:related} we present the related works. In Section \ref{section:experiments} we describe the details about measurement experiment set-up.
Then in Section \ref{section:results} we present and discuss the measurement results. We conclude and expand our future work in section \ref{section:conclusion}.
\section{Related Works}
\label{section:related}
P2P applications are quickly becoming one of the most important traffic generators  on the Internet \cite{pam05:ft}.
Consequently, there have been  lots of measurement studies about P2P systems to understand the impact on the traffic on the Internet, but a few measurement studies of P2P live streaming systems have been conducted.\\
In P2P file sharing applications measurements, 
Liang and al \cite{compnet06:kazaa} inferred the overlay building mechanisms of Kazaa.
For BitTorrent, Legout and al \cite{imc06:enoughchoke} made active measurements to evaluate the performances of the BitTorrent's algorithms, while Izal \cite{pam04:dissecting} or Pouwelse \cite{iptps05:bitmeasure} 
asses these algorithms too. 
Motivated by BitTorrent model, Guo and al. \cite{imc05:bittorrent} analyzed representative Bittorrent traffic to build a graph based multi-torrent model to study inter-torrent collaboration.\\
Nowadays, P2P VoIP systems like Skype are attracting more researchers.
Baset and al. \cite{infocom06:skype} show some of Skype's implementation details since it is a proprietary application, whereas Suh and al. \cite{infocom06:relayedskype} aim to characterize
 the nature of the skype relayed traffic and Guha and al. \cite{iptps06:skype} argue that the Skype traffic differs significantly from traffic in P2P file sharing networks.\\ 
Measuring P2P live streaming systems is still an emerging topic, but there are previous measurement studies about live streaming media delivered on the Internet.
As an important result, Sripanidkulchai and al.\cite{sigcomm04:feasibility} show that large-scale live streaming can be supported by P2P end-users applications despite the heterogenous capacity of peers.
In P2P IPTV systems, Zhang and al. \cite{invited:coolstreaming} present their own P2P IPTV system 
and give network statisctics like users' behavior in the whole system  and the quality of video reception.
Liang and al. \cite{wiptv06:pplive} have a similar work to ours. They study an existing P2P IPTV application by collecting packet traces.
Our work is different from theirs since we do not focus on a single application but on several representative applications to make comparisons between them. 
It helps us to highlight design differences and to infer global behavior of such P2P network without being strongly related to a single P2P system implementation.
We collected a larger and more various data set from an exhaustive panel of applications during an entire large-scale event. 
Finally, an important distinction between Liang works and ours come from the live interest of the measured event. 
It is intuitive but corroborated by Veloso and al \cite{imw02:live} that traffic patterns have not the same characteristics whether it exhibits a live interest or not.
\section{Experiments}
\label{section:experiments}
\begin{figure}[!ht]
\centering
\includegraphics[height=3.00 cm]{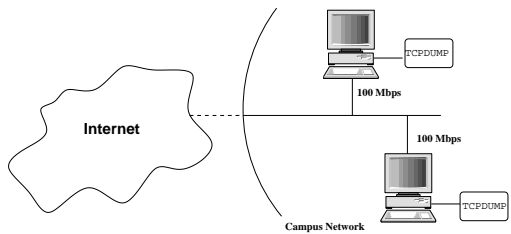}
\caption{Measurement experiments Platform. Each node is a common PC directly connected to the Internetvia campus network}
\label{fig:exp}
\end{figure}
Our measurement started with  the FIFA world cup from June 2nd to July 9th. 
We collected a huge amount of data, measuring most of the world cup games with different applications at the same time, under different network environnement (Ethernet access or residential ADSL access).
In this paper we focus on comparisons between four P2P IPTV applications according to their traffic pattern. 
In all our data, we selected packet traces on June 30.
Two soccer games were scheduled: one in the afternoon (Germany vs. Argentine) and one in the evening (Italy vs. Ukraine).
Figure~\ref{fig:exp} describes our measurement experiment platform.
We used two personal computers (PCs) with 1,8 GHz CPU, and common graphic capabilities. For the rest of this paper, the PCs will be called \emph{nodes}.
Operating system was \texttt{Windows XP} because all the applications have been implemented for this OS.
The two nodes were situated in our campus network and were directly connected to the Internet with 100 Mbps \texttt{Ethernet} access. 
We used \texttt{tcpdump} to collect the packets and their payload generated by the applications.
During each game, the nodes were running \texttt{tcpdump} and a distinct P2P application.
All the measurements have been done watching CCTV5, a Chineese sport channel.
We did not measure the traffic between the two games.
From the first game to the second one, we only changed the running P2P applications on the nodes. 
At the end of the experiments, we collected four packet traces: one per application.
The first game was measured by running PPStream and SOPcast on the nodes, and the second one by running PPLive and TVants.
Table \ref{tab:summary} summarizes the four collected traces. 
\begin{table}
\begin{scriptsize}
\caption{Packet traces summary}
\label{tab:summary}
\begin{center}
\begin{tabular}{|c|c|c|c|c|c|c|c|}
\hline
 & \texttt{Event} & \textbf{Trace size}  & \textbf{Duration} & \textbf{TCP UP} & \textbf{TCP DOWN} & \textbf{UDP UP} & \textbf{UDP DOWN} \\
 & & (MBytes) & (seconds) &(\%) & (\%) & (\%) & (\%) \\ \hline
\texttt{PPStream} & game 1 & 4,121 &12,375 & 79.50\% & 20.50\% & none & none \\ \hline 
\texttt{SOPcast} & game 1 & 5,475  & 12,198 & 3.89\% & 0.23\% & 79.98\%  & 15.90\% \\ \hline 
\texttt{PPlive} & game 2  & 6,339 & 13,321  &  85.81\% & 14.09\% & 0.08\% & 0.02\% \\ \hline 
\texttt{TVants} & game 2 & 3,992  & 13,358  &  61.67\% & 14.71\% & 13.57\% & 10.05\% \\ \hline 
\end{tabular}
\end{center}
\end{scriptsize}
\end{table}
The two measured events are not exactly the same but they are similar (a soccer game in the FIFA world cup) and exhibit the same live interest for users.
We analyze our packet traces by developing our own \texttt{perl} parsing tools. \\
Our platform has high-speed access and our observations can not be directly generalized to residential peers with common access to the Internet.
However, residential network capacities are quickly increasing and will have such high-speed access in only a few years when P2P IPTV would be commonly used.
\subsection{Data Collection Methodology}
We differentiate TCP sessions according to TCP Flags and we only take a session into consideration if at least one of its TCP segment has a payload.
Session durations are driven by TCP segment payload.
A session start time was calculated as soon as we received (or send) a TCP segment with a data payload. 
The session duration was increased for each new TCP segment with a payload.
A session ended when we received an explicit flag, but the end session time was the instant where we received the last TCP segment with payload.
The session duration depends only on TCP segment with payload.
In the same way, we compute session duration relying on UDP in the same payload-driven way. 
\section{Results}
\label{section:results}
During the experiments, we could visually control video quality playback for  each application.
Most of the applications, except SOPcast, worked relatively well and just suffered brief video playback troubles.
Sopcast was suffering a black screen during several minutes and it will be observed in the following results.
\subsection{General Observations}
According to  table~\ref{tab:summary}, PPStream relies only on TCP.  Major part of PPLive traffic relies on TCP whereas SOPcast traffic relies mostly on UDP. 
TVants is more balanced between TCP and UDP. In the following, some precisions have to be assessed: a download peer sends data to our nodes and an upload peer receives data from our nodes.
Collected packets can only provide from our two nodes or from a remote peer in the Internet.
\begin{figure*}[!ht]
\begin{center}
\subfigure[PPStream]{
\includegraphics[width=2.75 cm]{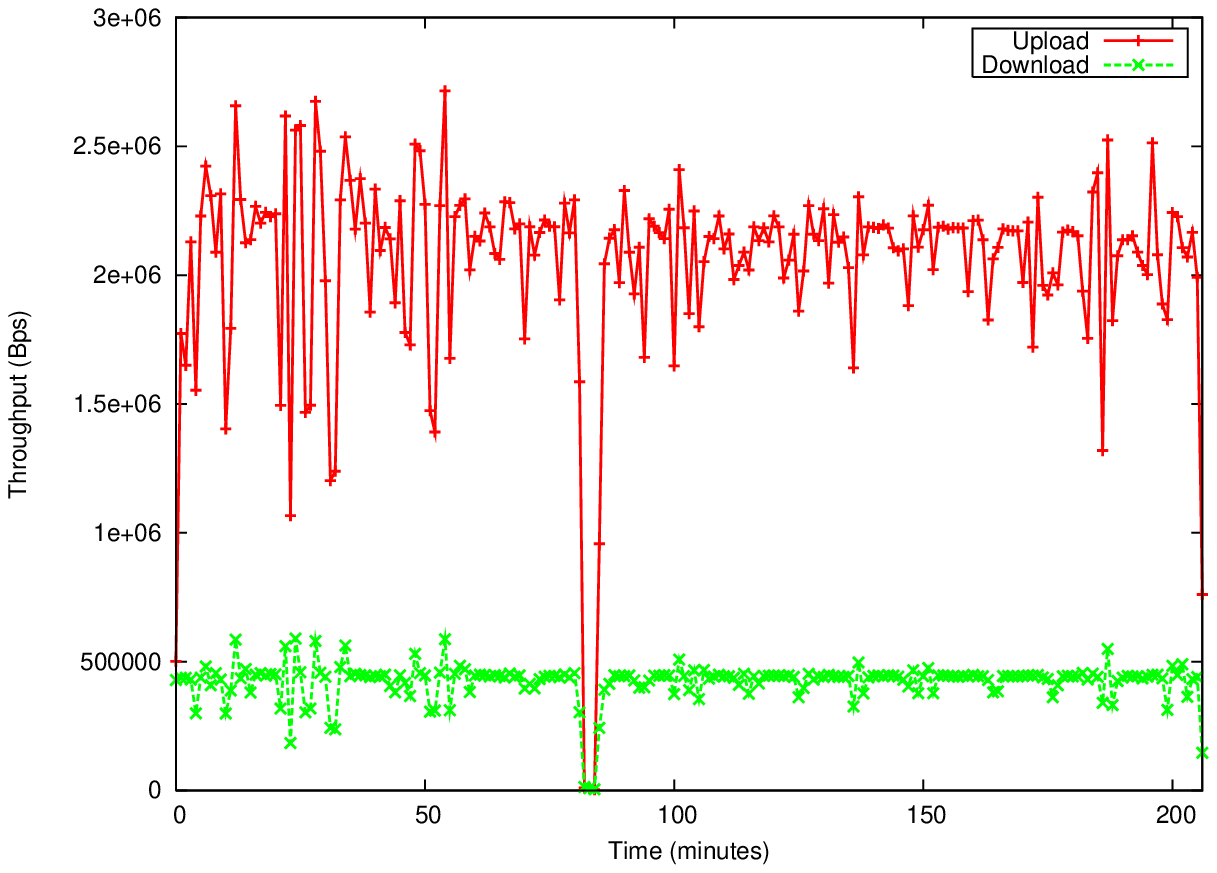}
\label{subfigure:throughput_pplive}
}
\subfigure[SOPcast]{
\includegraphics[width=2.75 cm]{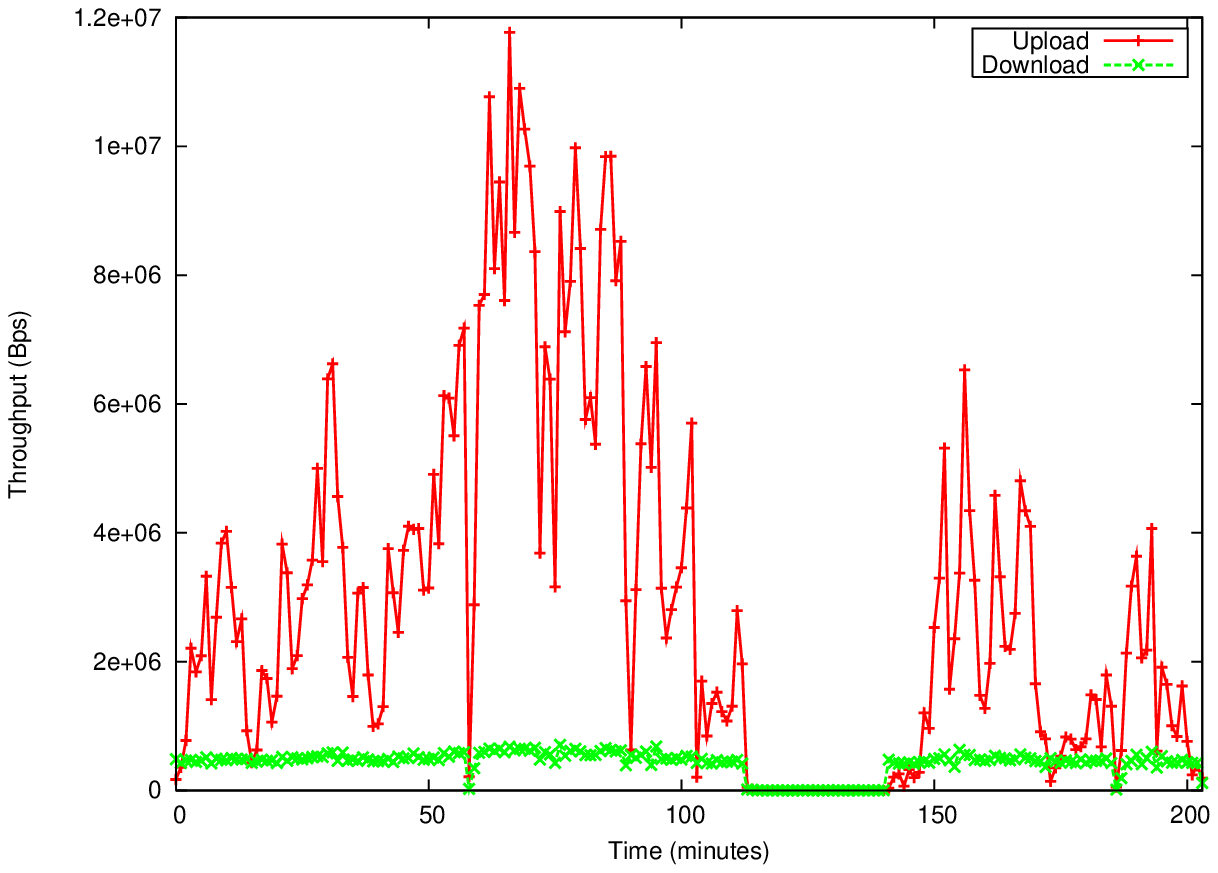}
\label{subfigure:throughput_ppstream}
}
\subfigure[PPLive]{
\includegraphics[width=2.75 cm]{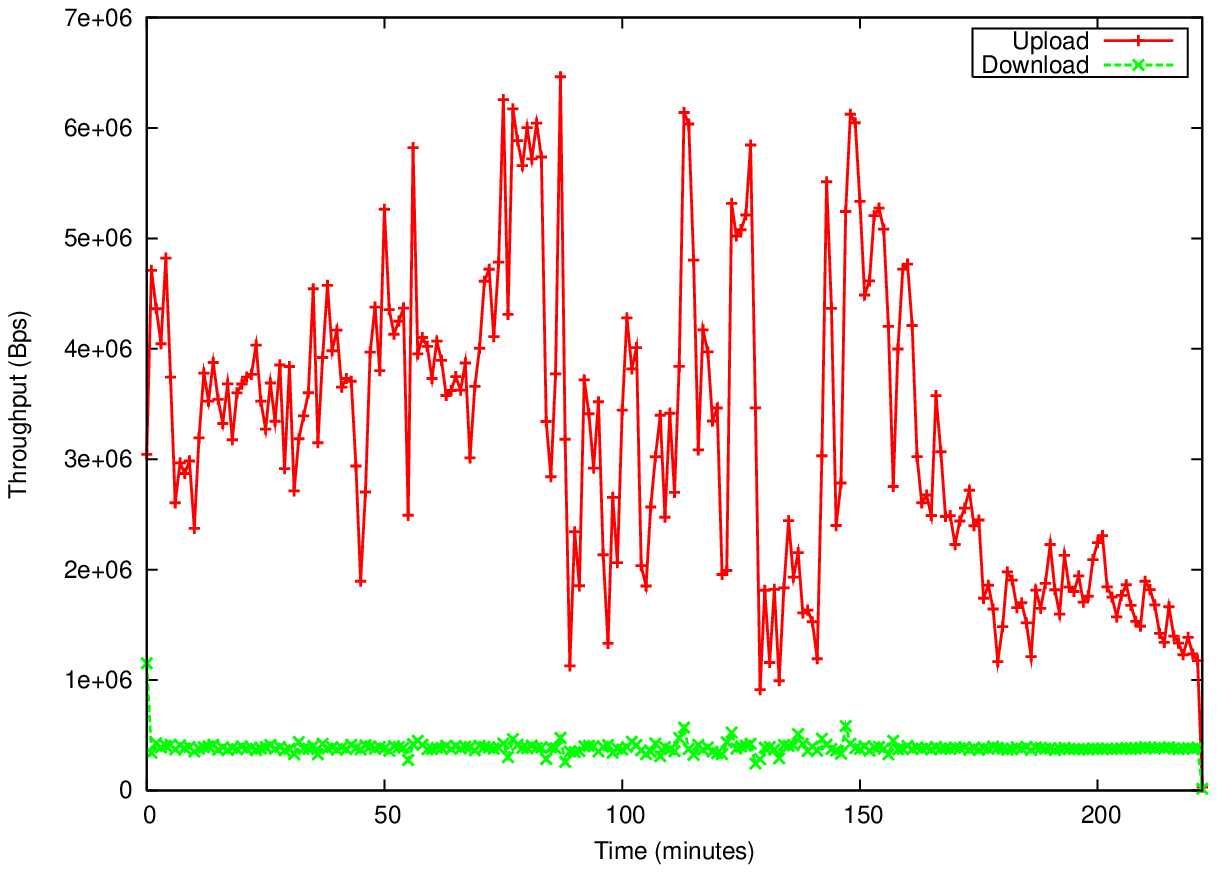}
\label{subfigure:throughput_sopcast}
}
\subfigure[TVants]{
\includegraphics[width=2.75 cm]{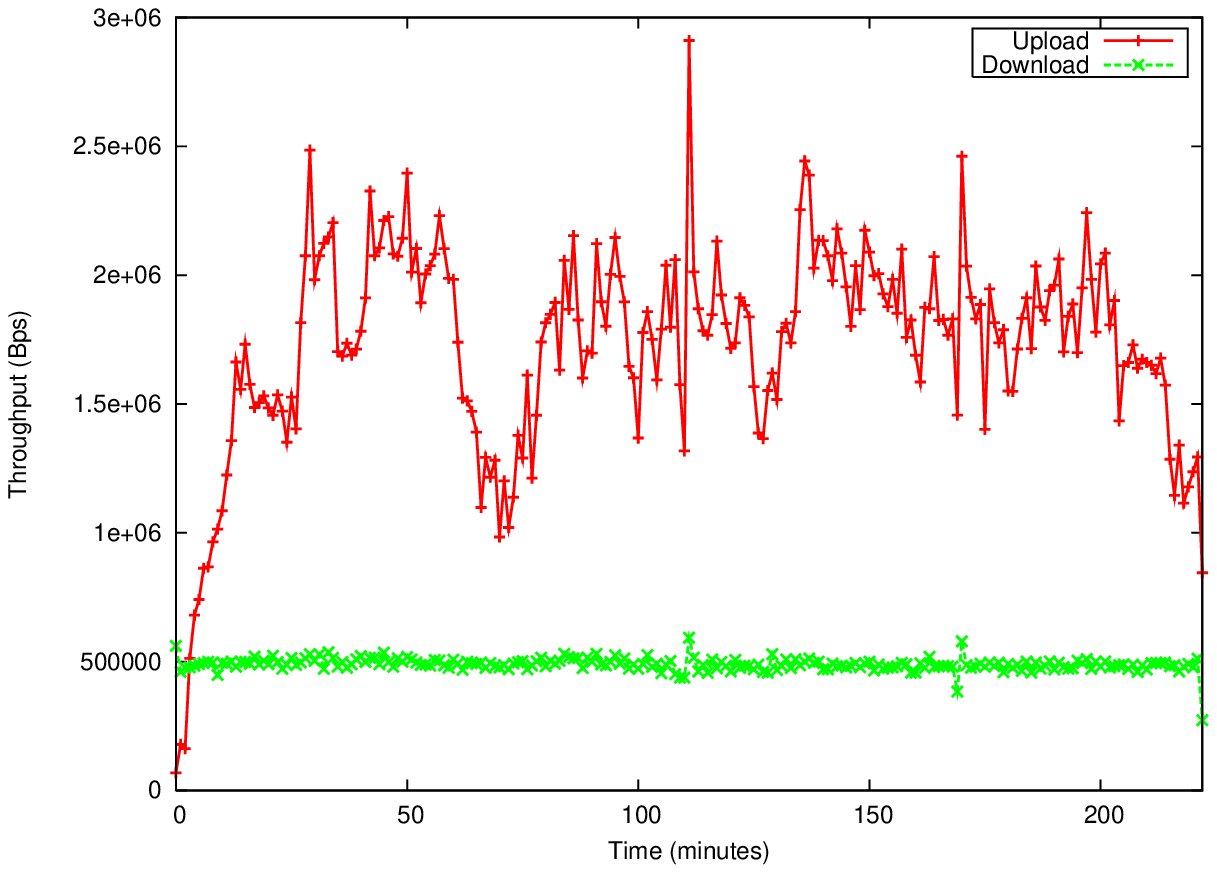}
\label{subfigure:throughput_tvants}
}
\caption{Total download (green curves) and upload (red curves) throughput. Bin duration is 60 seconds}
\label{fig:throughput}
\end{center}
\end{figure*}
Figure~\ref{fig:throughput} shows the total download and upload throughput for all the applications.
Download throughput is quite constant while upload fluctuates largely and at a higher rate.
This was expected because the nodes attempt to download video at a constant bitrate and they have wide upload capacities.
Sopcast (Fig.~\ref{subfigure:throughput_sopcast}) had a very low traffic from minutes 130 to minutes 140 and we  watched a black screen during the period. 
The problem does not occur for network reasons because PPStream was working well during the same period.
The video source has probably suffered technical problem during this period.
\subsection{Traffic Patterns}
These P2P applications are proprietary but claim to have swarming mechanisms where peers exchange information about data chunks and peers (\cite{infocom05:donet}).
In such P2P network, a peer will iteratively discover other peers and would establish new signaling or video sessions.
Video sessions are likely to have long duration because users want to watch the entire game whereas signaling sessions are likely to be shorter in time.
Furthermore, video streaming packets size is expected to be large and signaling session packets size is supposed to be common.
\begin{figure*}[!ht]
\begin{center}
\subfigure[PPStream]{
\includegraphics[width=5.50 cm]{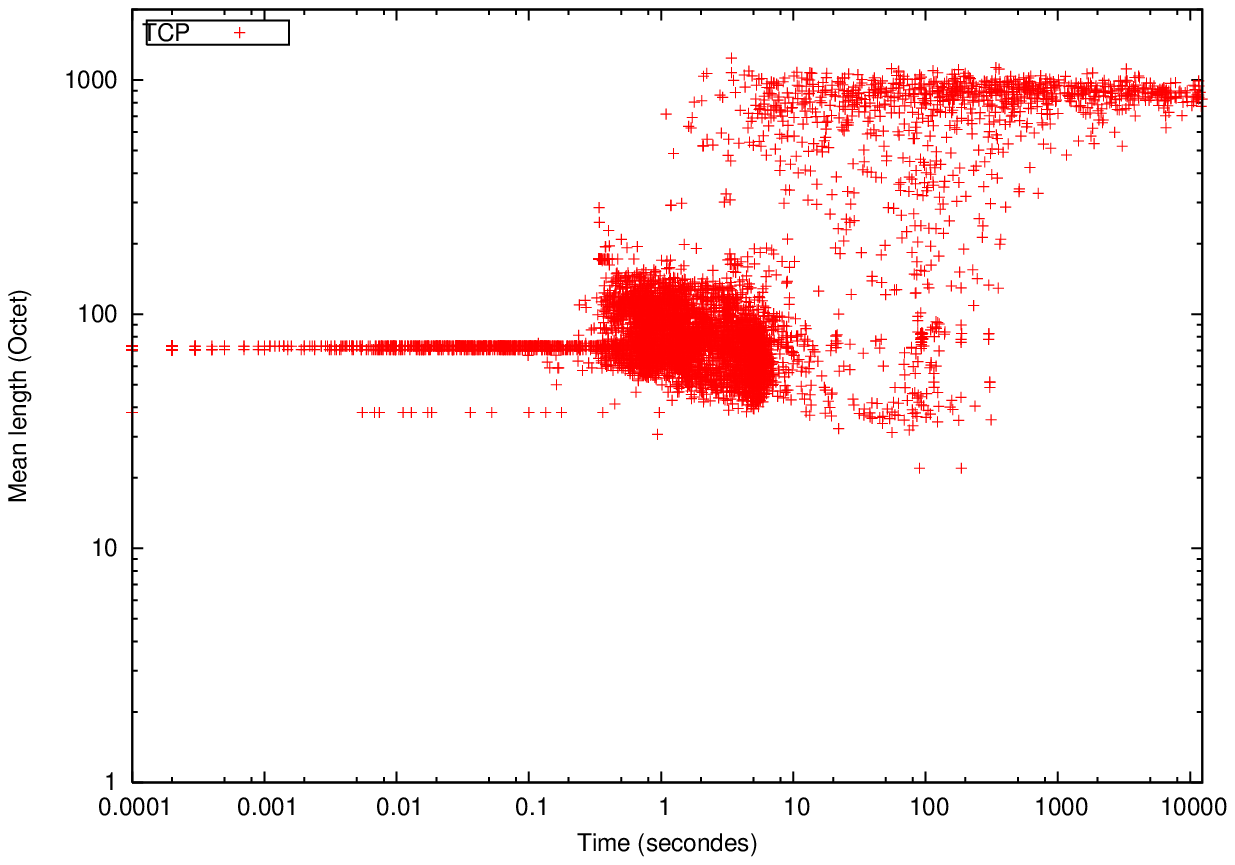}
\label{subfigure:nuage_ppstream}
}
\subfigure[SOPcast]{
\includegraphics[width=5.50 cm]{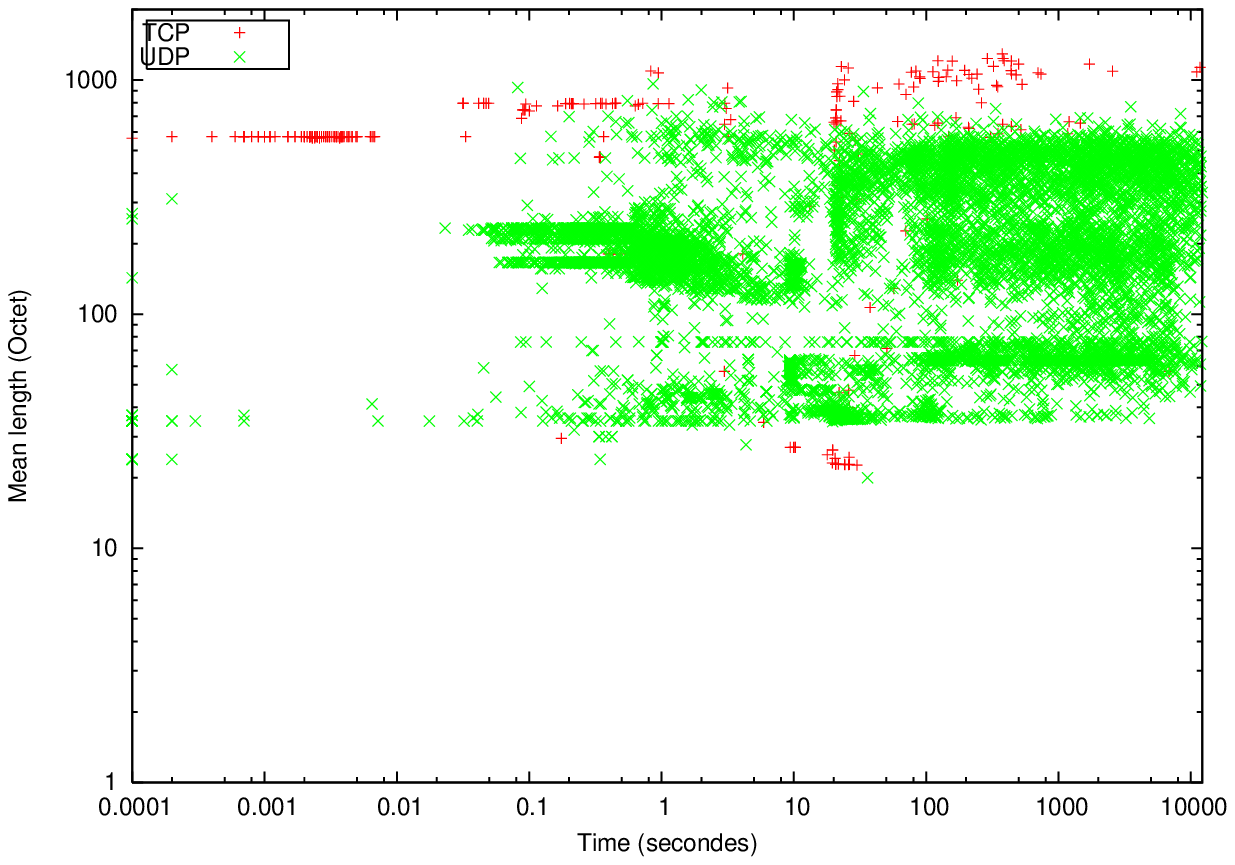}
\label{subfigure:nuage_sopcast}
}
\subfigure[PPLive]{
\includegraphics[width=5.50 cm]{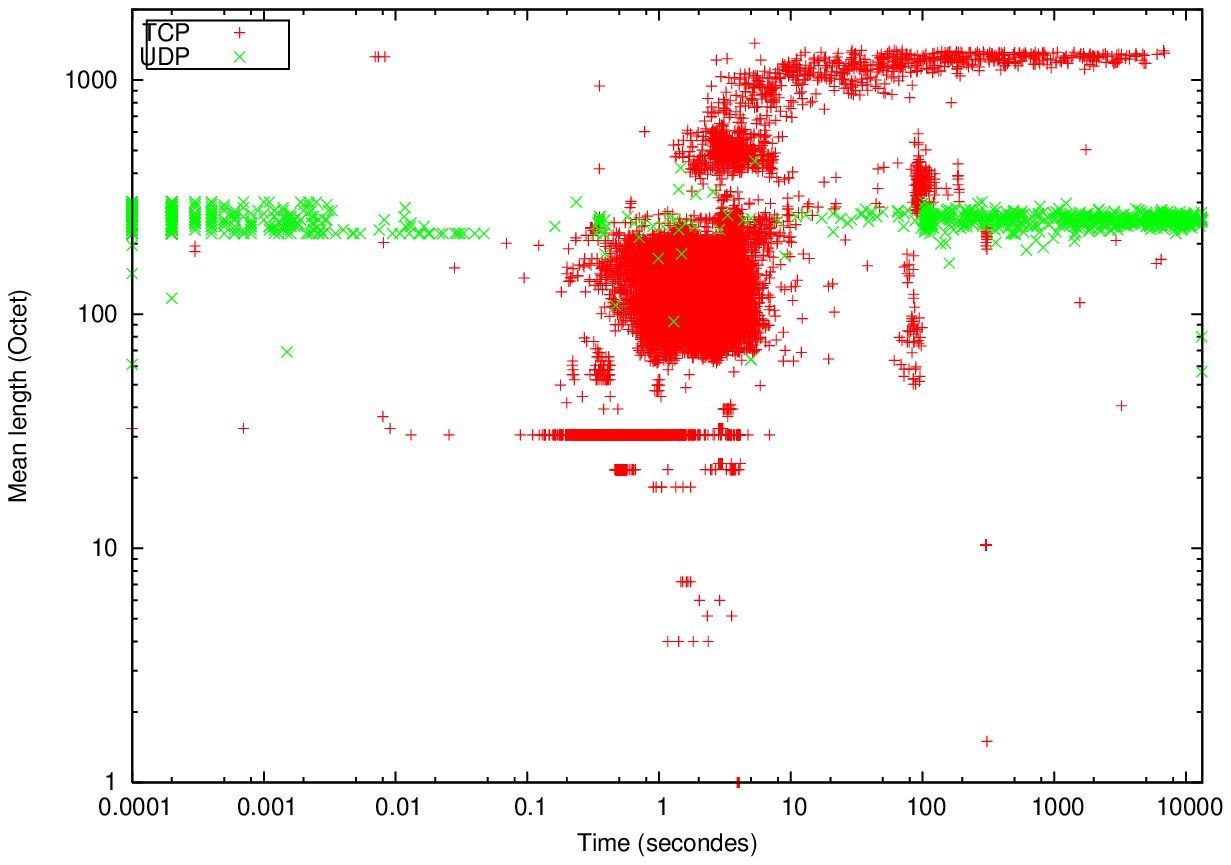}
\label{subfigure:nuage_pplive}
}
\subfigure[TVants]{
\includegraphics[width=5.50 cm]{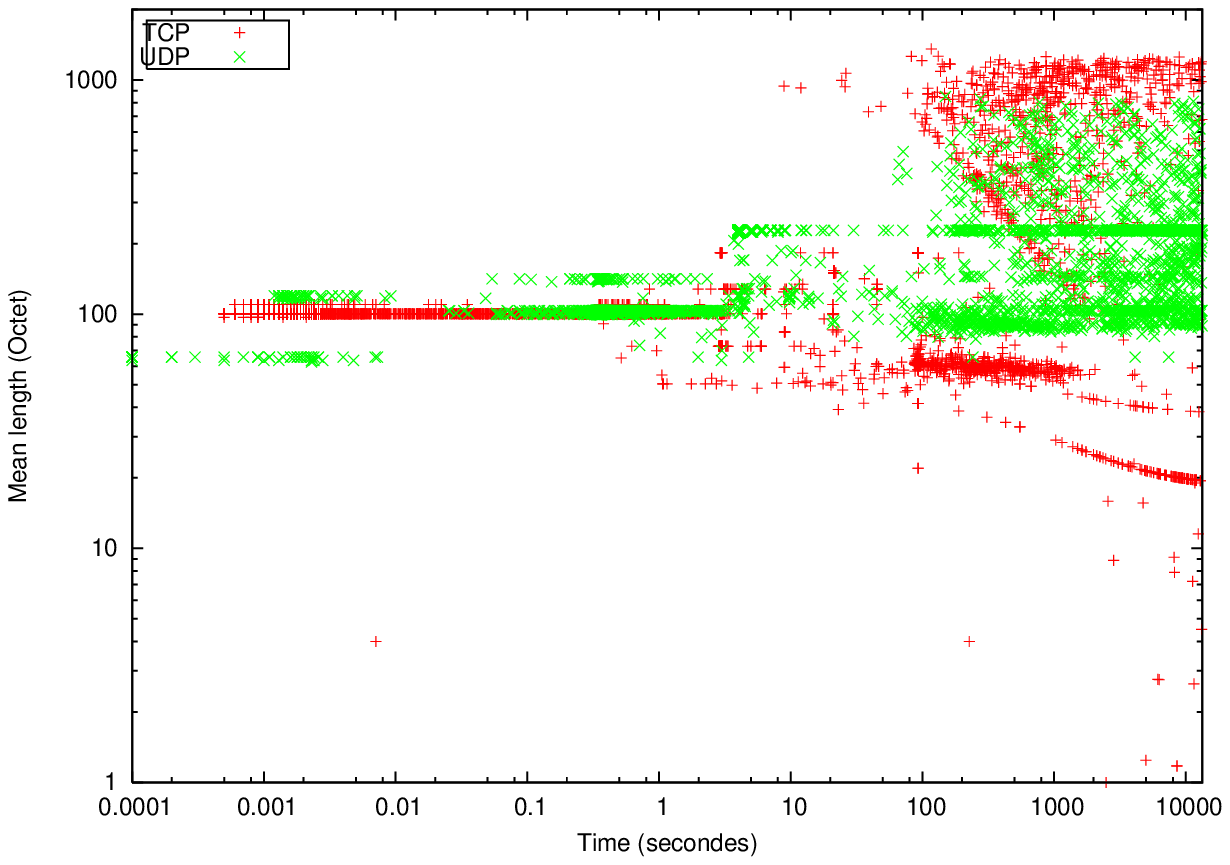}
\label{subfigure:nuage_tvants}
}
\caption{Average packet size according to peers session duration}
\label{fig:nuage}
\end{center}
\end{figure*}
Figure~\ref{fig:nuage} plots the average packet size according to the session duration.
PPLive (Fig.~\ref{subfigure:nuage_pplive}) and PPStream (Fig.~\ref{subfigure:nuage_ppstream}) have the same pattern if you except that PPLive uses UDP. 
PPLive UDP session can have long lifetime, but the average packet size is small and constant.
PPLive UDP traffic transports certainly signaling traffic. 
PPLive and PPStream exhibit two clusters in their traffic pattern: 
the one in the middle of the plot is for signaling session (small packet and short session duration) and the one in the right top of the plot is for video session (large packet and long session).
SOPcast (Fig.~\ref{subfigure:nuage_sopcast}) and TVants (Fig.~\ref{subfigure:nuage_tvants}) patterns still exist but are not so clearly formed. 
PPLive and PPStream trasnport necessarily video on TCP. SOPcast should mostly used UDP but there is maybe a few video sessions relying on TCP. According to its video cluster (Fig.~\ref{subfigure:nuage_tvants}), 
TVants should transport video on both TCP or UDP.
We separated video and signaling traffic with an already proposed heuristic (\cite{wiptv06:pplive}).
The heuristic works as follows: for a session (same IP addresses and ports), we counted the number of packet bigger or equal than 1000 Bytes\footnote{We chose 1000 Bytes instead of 1200  because we think it would fit better our traces}.
If a session had at least 10 large packets, then it was a video session and we removed non-video packet from the session. 
%
%Using the heuristic for the four traces, we found that video traffic represent about 97\% of all the traffic (PPStream 99\%, SOPcast 97\%, PPLive 94\% and TVants 97\%).
Using the heuristic for the four traces, we found that video traffic represents about 88\% of the overall traffic (PPStream 86\%, SOPcast 80\%, PPLive 96\% and TVants 90\%).
\subsection{Peers Download Policies}
\begin{figure*}[!ht]
\begin{center}
\subfigure[PPStream]{
\includegraphics[width=5.50 cm]{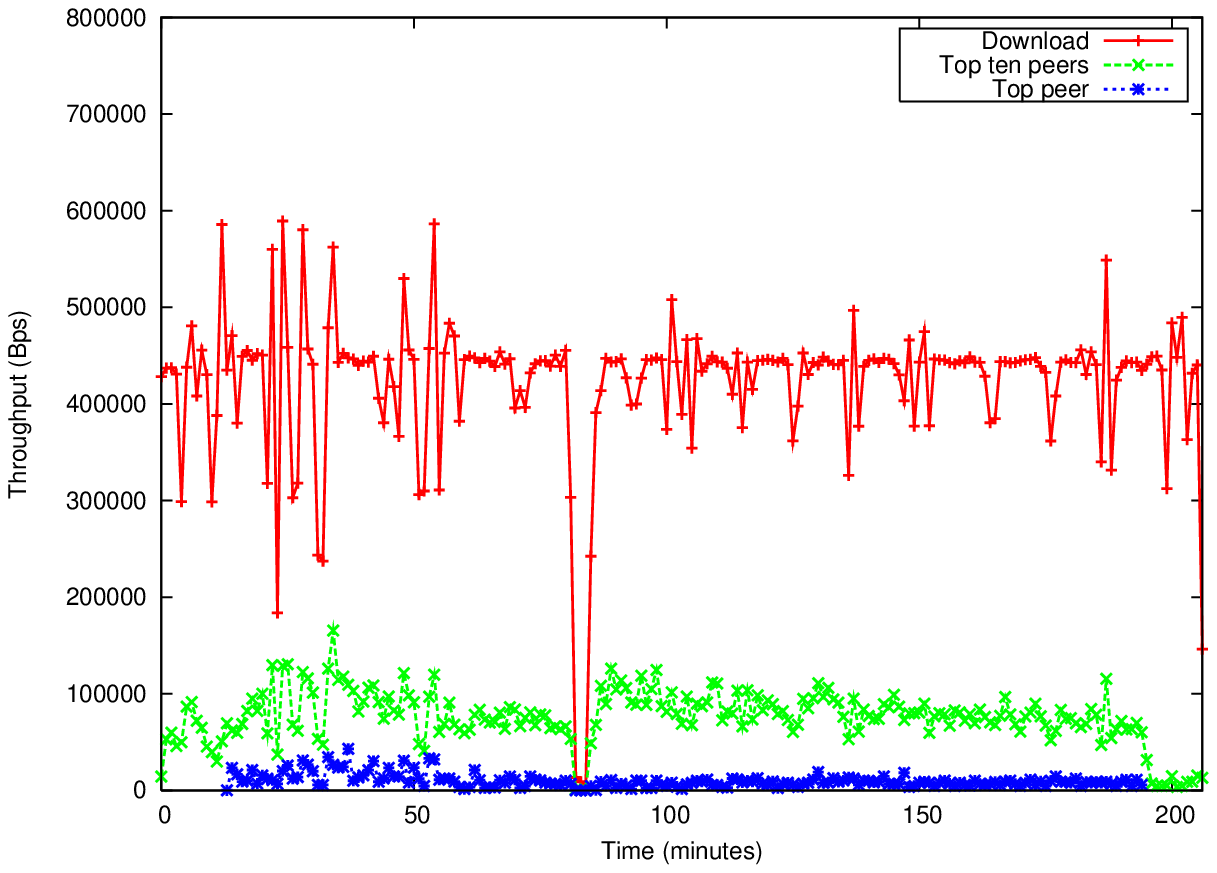}
\label{subfigure:download_ppstream}
}
\subfigure[SOPcast]{
\includegraphics[width=5.50 cm]{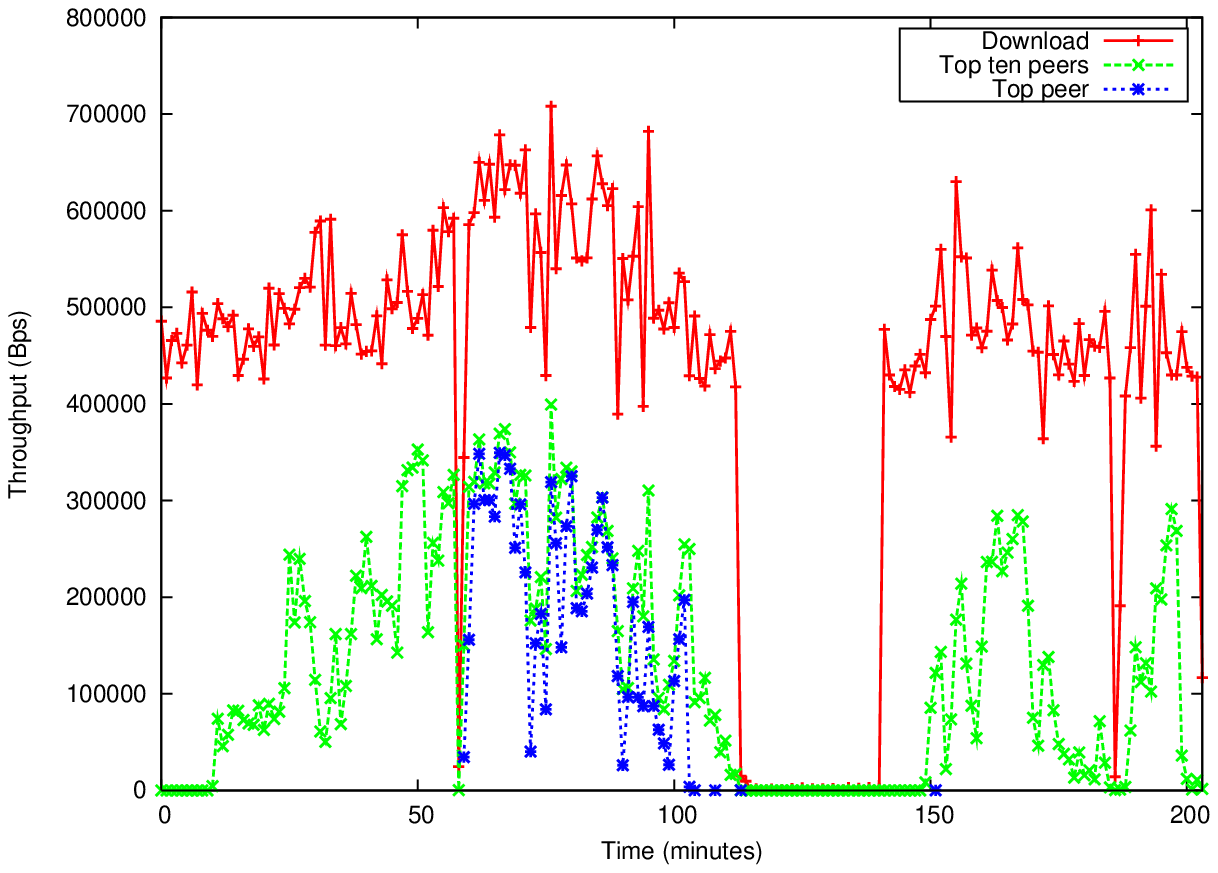}
\label{subfigure:download_sopcast}
}
\subfigure[PPLive]{
\includegraphics[width=5.50 cm]{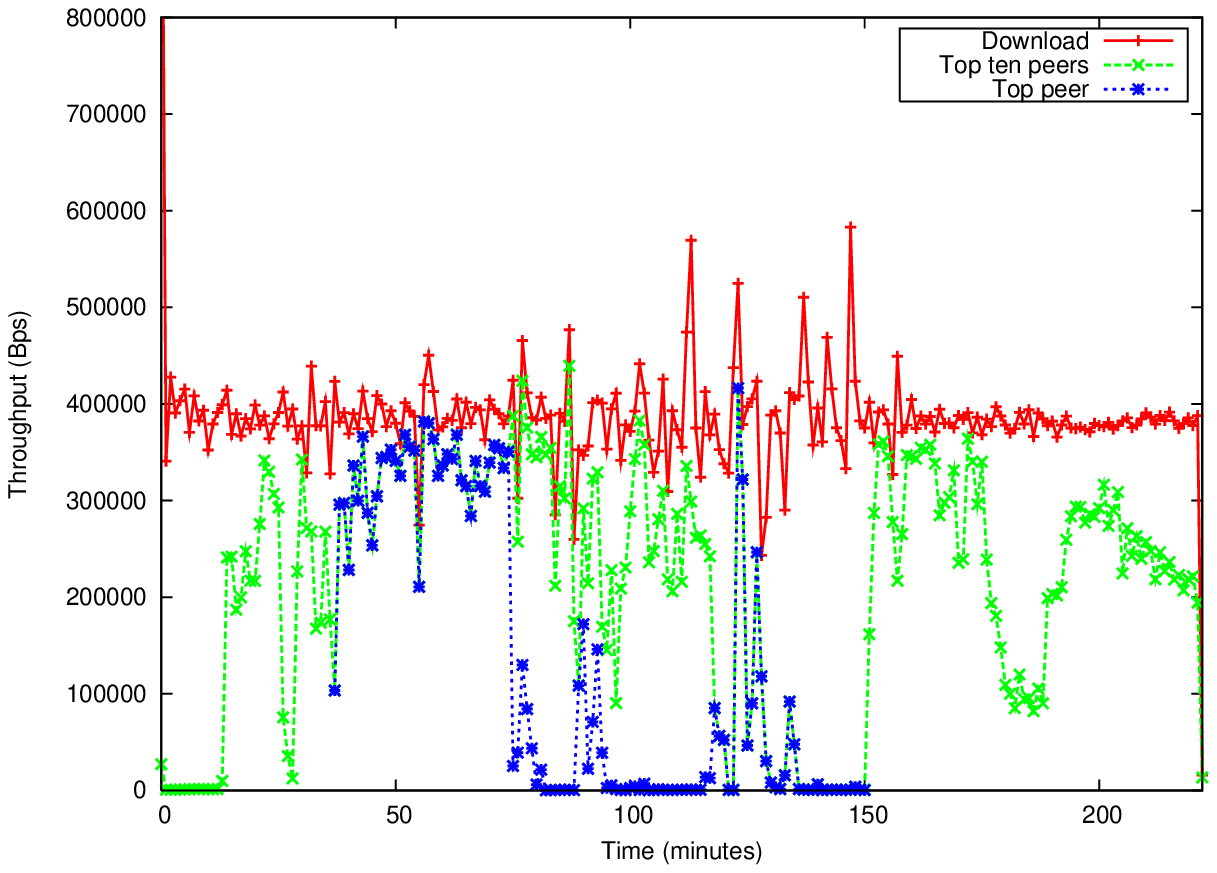}
\label{subfigure:download_pplive}
}
\subfigure[TVants]{
\includegraphics[width=5.50 cm]{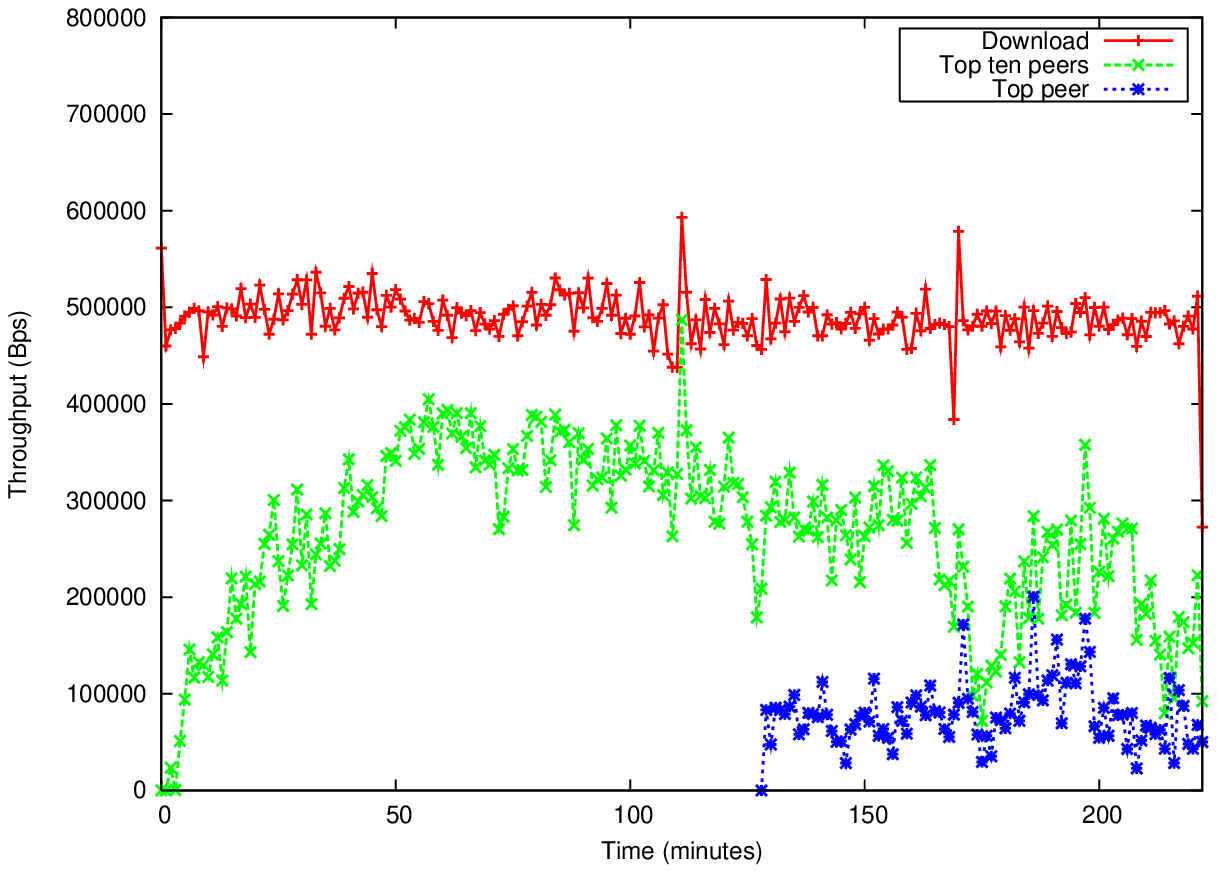}
\label{subfigure:download_tvants}
}
\caption{Download: Total traffic, top ten peers traffic and top peer traffic. Bin duration is 60 seconds}
\label{fig:download}
\end{center}
\vspace{-0.5cm}
\end{figure*}
We wanted to understand how our nodes download the video among the other peers.
For each trace, we computed the amount of data that our nodes downloaded from each of the other  peers.
We isolated the ten top peers (peers which sent the biggest amount of data to our nodes across the entire traces duration). We isolated the top peer in the same way (top peer belongs to top ten peers).
In figures~\ref{fig:download}, each plotted value is a 60 seconds average interval (bin duration is 60s).
The download policy of PPLive (Fig.~\ref{subfigure:download_pplive}) seems to periodically switch from a peer to another one. During its session duration, the top peer contributes to the  major part of the download traffic.
PPStream download policy (Fig.~\ref{subfigure:download_ppstream}) is different: the top ten peers do not contribute to a large part of the download traffic.
PPStream seems to get the data from many peers at the same time and its peers seem to have long session duration.
PPLive seems to get the data from only a few peers at the same time but its peers have not a long session duration.
Sopcast top ten peers (Fig.~\ref{subfigure:download_sopcast}) contribute to about half the total download traffic and top peer could contribute to half the total traffic during its lifetime too.
In a way, Sopcast download policy looks like PPLive policy: it switches  periodically from provider peer. However, SOPcast seems to always need more than a peer to get the video
 compare to PPLive where a single peer could be the only video provider.
TVants download policy (Fig.~\ref{subfigure:download_tvants}) seems to mix PPStream and SOPcast policies. TVants top ten peers contribute to about half the total download traffic (SOPcast), but top peer does not contribute to a large amount of the total traffic (PPStream).
TVants top peer do not contribute as few as PPStream's one but do not stay as long as PPStream top peer.
If we summarize our observations,  the download policies of all the applications are different and do not need the same peer capabilities.
Some policies expect peers to stay in the network for a long time (PPStream) or short time (PPLive), or expect peer to have huge capacities (PPLive) or low (PPStream). Some polices seems to mix other policies (SOPcast, TVants).
\subsection{Peers Behavior}
\begin{figure}[!ht]
\centering
\includegraphics[height=5.50 cm]{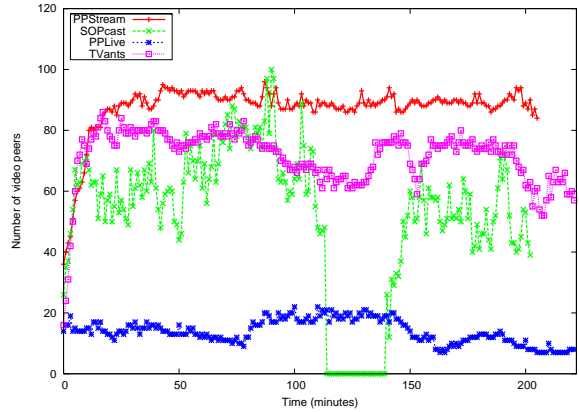}
\caption{Churn of download Peers. Bin duration is 60 seconds}
\label{fig:churn}
\vspace{-0.5cm}
\end{figure}
\begin{figure*}[!ht]
\begin{center}
\subfigure[PPStream. Average lifetime = 1222s]{
\includegraphics[width=5.50 cm]{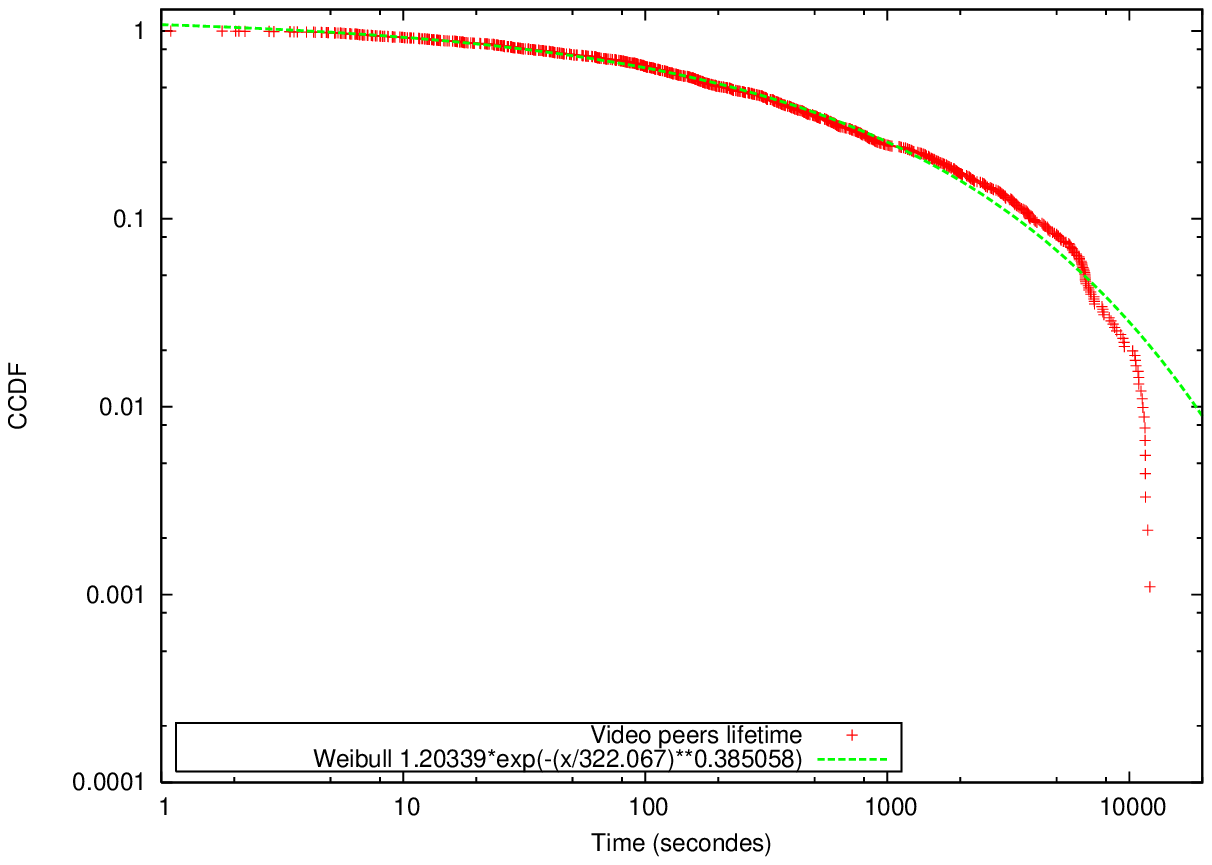}
\label{subfigure:lifetime_pplive}
}
\subfigure[SOPcast. Average lifetime = 1861s]{
\includegraphics[width=5.50 cm]{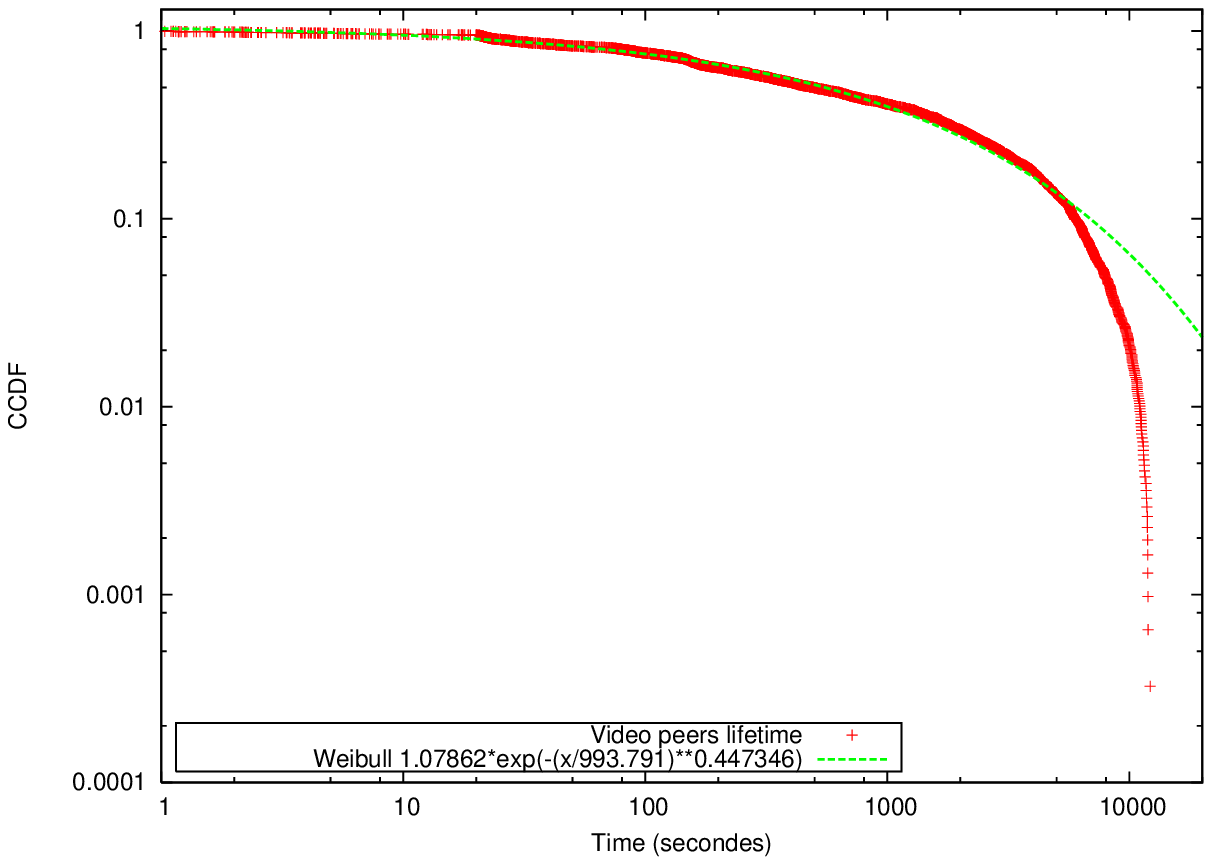}
\label{subfigure:lifetime_ppstream}
}
\subfigure[PPLive. Average lifetime = 393s]{
\includegraphics[width=5.50 cm]{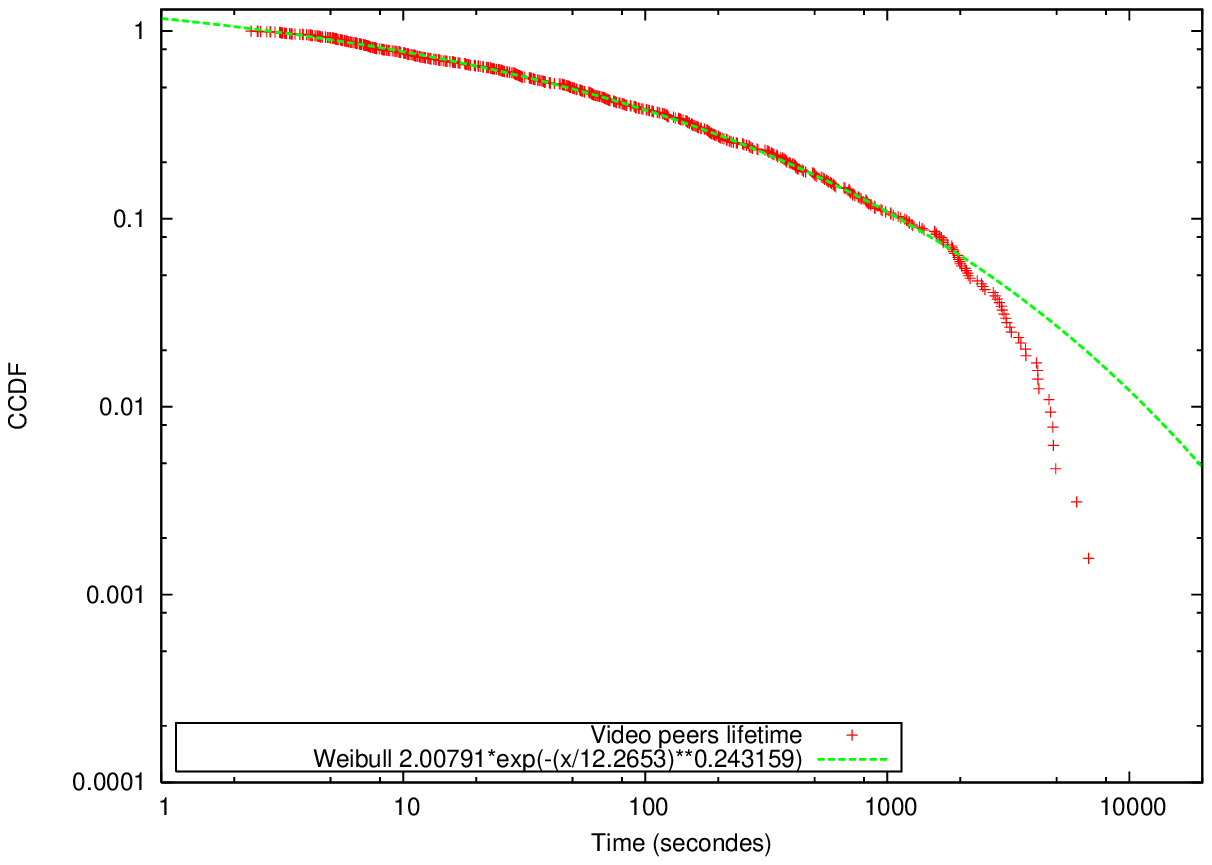}
\label{subfigure:lifetime_sopcast}
}
\subfigure[TVants. Average lifetime = 2778s]{
\includegraphics[width=5.50 cm]{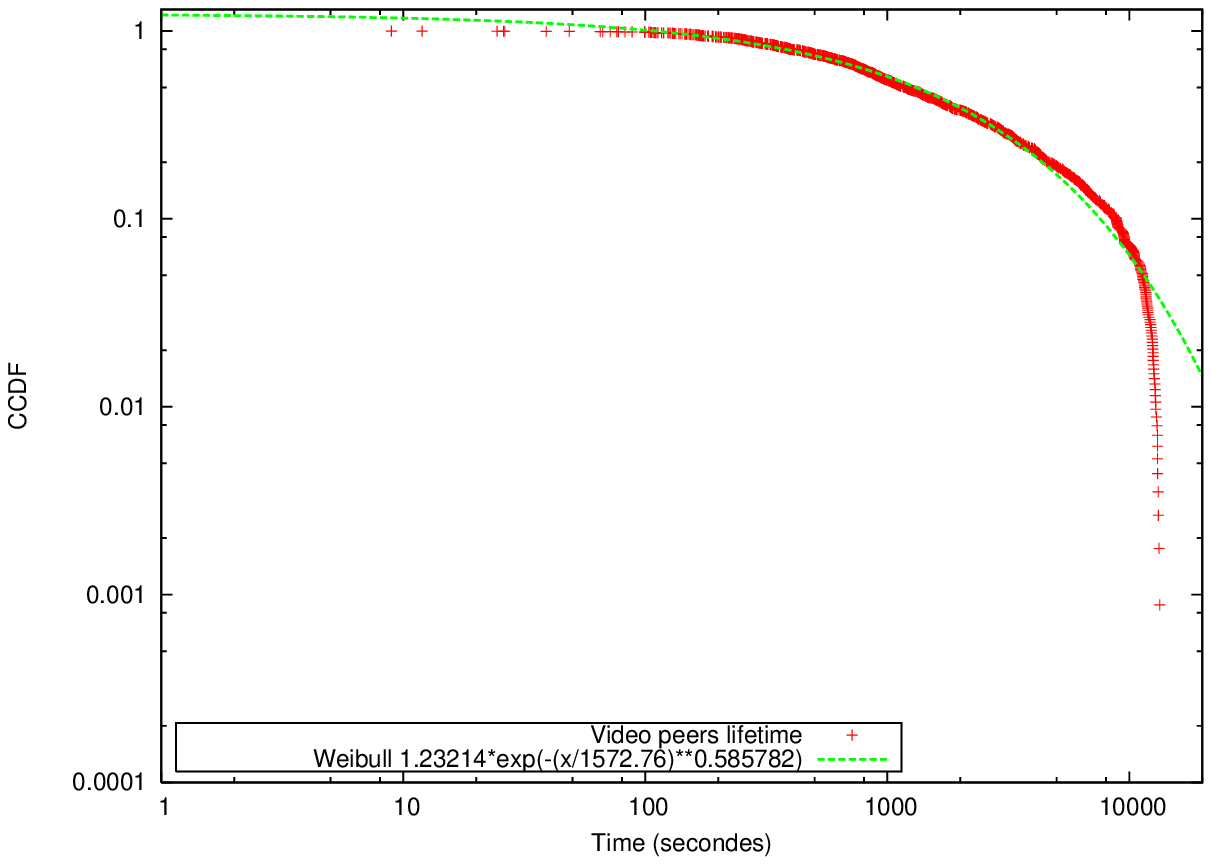}
\label{subfigure:lifetime_tvants}
}
\caption{Video peers lifetime}
\label{fig:lifetime}
\end{center}
\vspace{-1cm}
\end{figure*}
P2P networks have to deal with the arrivals and departures of peers (churn of peers), and it is a more complicated problem with video live streaming because it is time sensitive. 
With figure~\ref{fig:churn}, the churn of peers exhibits different peer's behavior for each application. PPLive has a relatively low and constant number of video download peers whereas PPStream has
a high number.
SOPcast has as many peers as PPStream but its number of peers fluctuates. As expected, SOPcast has no peer when it receives no traffic.
PPLive and PPStream have their peers behavior which corroborates the traffic observation: a large set of steady peers for PPStream and only a reduced set for PPLive.
The set of Sopcast peers flutuates widely and so does its download traffic. As expected, SOPcast has no video peers when it receives no traffic. By the way, PPStream seems to suffer 
a download traffic collapse at minute 90 (Fig~\ref{subfigure:throughput_ppstream}) but we do not observe a decreasing number of peers at this time. 
This phenomena will be more precisely analyzed in our ongoing work.
TVants looks like both PPStream for the large set of peers and SOPcast for the fluctuations.
TVants used TCP and UDP whereas PPStream relies only on TCP and SOPcast mostly on UDP. Since peers churn in TVants seems to exhibit behavior from the two others, more investigation should be done
to correlate churn and transport protocol. 
The complementary cumulative distribution function (CCDF) of video sessions are plotted on figure~\ref{fig:lifetime}. All the distribution seems to be Weibull distribution. 
The measured events were soccer games, and for all the applications there is no more than  10\% of peers which stay in the network during an entire game (5400 seconds).
The average lifetime of the video session is quite different according to the application. PPLive average lifetime (Fig.~\ref{subfigure:lifetime_pplive}) is shorter than all the other applications.
A previous intuition was that PPlive switches voluntarily between peers to get the video. PPLive average lifetime could eventually be explained by its design and not by a particular user behavior. 
\section{Conclusion}
\label{section:conclusion}
In this work, we study P2P IPTV network traffic during moments of a large-scale event. 
During a same event, the amount of exchanged data for all the applications is quite similar.
In this paper, our analyzes highlight for the application different traffic patterns and underlying mechanisms.
Each application does not manage its P2P networks exactly in the same way, since we show download policies, churn of peers and session durations are different.
With regards to all the data we collected, our results are promising but the observations have to be generalized.
At the present time, we are analyzing packet traffic of the same applications during other games and under different network accesses (corporate and residential high-speed accesses). 
Our next results would help us to corroborate and clarify our observations. 
%
%\bibliographystyle{IEEEtran}
%\bibliography{biblio}

%
\end{document}